\documentclass[aps,prd,showpacs,nofootinbib,preprintnumbers,twocolumn]{revtex4-1}

\usepackage{bm}
\usepackage{latexsym}
\usepackage{natbib}
\usepackage{url}
\usepackage{dcolumn}
\usepackage{color}
\usepackage{amsfonts,amssymb,amsmath}
\usepackage{graphicx,epsfig}
\usepackage{psfrag}
\usepackage{subfigure}
\usepackage{natbib}

\begin{document}

\title{\textbf{Dark matter as scalaron  in $f(R)$ gravity models}}

%----------------------------------------------------------------------------------------------------------------------------------------------------------%

\author{ Bal Krishna Yadav}
\email[]{balkrishnalko@gmail.com}

\author{Murli Manohar Verma}
\email[]{sunilmmv@yahoo.com}

\affiliation{Department of Physics, University of Lucknow, Lucknow 226 007, India}
%--------------------------------------------------------------------------------------------------------------------------------------------------------%
\date{\today}

\begin{abstract}
We explore the scalar field  obtained under the conformal transformation of the  spacetime metric $g_{\mu\nu}$ from  the  Jordan frame to the  Einstein frame in $f(R)$ gravity. This scalar field is the result of the modification in the gravitational part of  the   Einstein's general relativistic theory of gravity. For $f(R)=\frac{R^{1+\delta}}{R_{c}^{\delta}}$, we find  the effective  potential of the scalar field and  calculate the mass of the scalar field particle ``scalaron".  It is shown that the mass of the scalaron depends upon the energy density of standard matter in the background (in solar system, $m_{\phi}\sim 10^{-16}$ eV).  The interaction between standard matter  and scalaron is weak in the high curvature regime. This linkage between the mass of the scalaron and the background  leads to the physical effects  of dark matter  and is expected to reflect   the anisotropic propagation of scalaron in moving baryonic  matter fields as in merging  clusters (Bullet cluster, the Abell 520 system, MACS etc.). Such scenario also  satisfies the local gravity constraints of $f(R)$ gravity. We  further calculate the equation of state of the scalar field in the action-angle variable formalism and show its distinct   features as the dark matter and dark energy  with respect to energy density of the scalar field at different values of the model parameter  $\delta$.
\end{abstract}
\maketitle
\section{\label{1}\textbf{Introduction}}
The explanation of the late-time accelerated expansion of the universe and dark matter  are the two major challenges of the present day cosmology. There are several  observational evidences such as   Supernovae Ia, Baryon Acoustic Oscillation (BAO), Cosmic Microwave Background anisotropies, weak gravitational lensing  etc. \cite{b1,b2,b3,b4} which indicate that the present universe is in the phase of accelerated expansion.
 There are several approaches to explain the late-time accelerated expansion and dark matter both. The most suitable model to explain such acceleration is the  Lambda Cold Dark Matter ($\Lambda$CDM)  model, where $\Lambda$ is the well known cosmological constant  \cite{b5,b6}. However, the cosmological constant faces two serious problems $(i)$ coincidence problem $(ii)$ fine tuning problem. The fundamental nature of the cosmological constant is unknown. An explicit matter component with strange characteristics is also introduced to explain this phenomena. It is known as the dark energy \cite{b7,b8,b9,b10}. There is still another approach called as modified gravity, wherein the late-time accelerated expansion is explained without using any explicit dark energy component \cite{b11,b12,b13,b14,c1}. In modified gravity models, we have $f(R)$ theory, Braneworld models,  Gauss-Bonnet dark energy models etc., out of which $f(R)$ theory is one of the simplest modified gravity models,  wherein   the Ricci scalar $R$ of  the  Hilbert-Einstein action is generalised into a function $f(R)$ of $R$. Besides the existence of the late-time accelerated expansion of the universe, the observations of rotation curves of galaxies \cite{b15,b16} and gravitational lensing indicate the presence of new matter often known as dark matter.   Rather strangely, this matter does not  experience the electromagnetic interaction, even though it has the gravitational interaction with the normal matter and radiation. The fundamental nature of the dark matter is still  mysterious. There are some famous candidates of dark matter like weakly interacting massive particles (WIMPs) \cite{c2}. At the same time, there exist   several alternative  approaches to explain the effects of  dark matter in modified gravity theories \cite{b17, b171,  b18, b19}.

Here, we  discuss the dark matter problem in $f(R)$ gravity. There are three approaches to derive the field equations in $f(R)$ models $(i)$ metric formalism $(ii)$ Palatini formalism  and  $(iii)$ metric-affine formalism.  We use the metric formalism  in  $f(R)$ gravity and   dark matter is  explained without considering any extra matter component.  In particular,  we consider $f(R) = \frac{R^{1+\delta}}{R_{c} ^\delta}$   model to explain the  geometrical effects of dark matter.
Previously, dark matter and dark energy  problems have been studied with scalar field in Ref. \cite{b20,b21}. The oscillations of the scalar field have  been shown to contribute to dark energy  in Ref. \cite{b18, b20, b25, b28}.   We find out the relation between equation of state $w$ and energy density $\rho_{\phi}$  of the scalar field for different model parameters.

Thus, the  present paper is organised in five sections. Section II  contains the basic field equations of the $f(R)$ gravity and the conformal transformation from the Jordan frame to the  Einstein frame. In Section  III,  the chameleon mechanism is discussed in the framework of our model and the mass of the scalaron as the dark matter particle has been calculated showing its dependence on the background matter density.  Section IV  contains the discussion about the  equation of state of the scalar field in the formalism of action-angle variable for different model parameters. We conclude and discuss our results in Section V.
\section{\label{2}\textbf{Field equations of $f(R)$ gravity and conformal transformation}}
We consider the $4-$dimensional action of the $f(R)$ gravity model given by some general function $f(R)$ of the Ricci scalar as
\begin{eqnarray} \mathcal{A}= \frac{1}{2\kappa^{2}}\int d^{4}x \sqrt{-g} f(R) + \mathcal{A}_{m}(g_{\mu\nu}, \Psi_{m})\label{a1}\end{eqnarray}
where $\kappa^{2}=8\pi G$ and $\mathcal{A}_{m}$ is the action of the matter part with matter field $\Psi_{m}$.
We assume  the spacetime as homogeneous, isotropic and spatially flat. It is given by the Friedmann-Lemaitre-Robertson-Walker (FLRW) spacetime as
\begin{eqnarray}ds^{2}= -dt^{2} + a^2(t)[dr^2 + r^2 (d\theta^{2} + \sin^{2}\theta d\phi^{2})]\label{a242} \end{eqnarray}
where $a(t)$ is time dependent scale factor and the speed of light $c=1$.

Here, we use the metric formalism in which connections $\Gamma^{\alpha}_{\beta\gamma}$ are defined in terms of the metric tensor $g_{\mu\nu}$. Varying the action  (\ref{a1})    with respect to  $g_{\mu\nu}$, we obtain  the field equations given by
\begin{eqnarray}F(R)R_{\mu\nu} - \frac{1}{2} f(R)g_{\mu\nu}\nonumber\\ - \nabla_\mu\nabla\nu F(R) + g_{\mu\nu}\Box F(R)=\kappa^{2}T_{\mu\nu},\label{a2}\end{eqnarray}
where $F(R)\equiv\frac{\partial{f}}{\partial{R}}$ and $T_{\mu\nu}$ is the  energy-momentum tensor for matter. The trace of field equations ($\ref{a2}$) is given by
\begin{eqnarray}   3 \Box F(R) + F(R) R -2f(R) = \kappa^{2} T, \label{a3}  \end{eqnarray}
where $T=-\rho + 3p$.  Here,  $\rho$ and $p$ are the energy density and  pressure of matter,  respectively.  The  trace  of  the field equations shows that the  Ricci scalar is dynamical if $f(R) \neq R$.
We rewrite the action (\ref{a1}) in the form
\begin{eqnarray}\mathcal{A} = \int\sqrt{-g}\left(\frac{1}{2\kappa^{2}}F(R)R - U \right)d^{4}x + \mathcal{A}_{m},\label{a4}                                \end{eqnarray}
where \begin{eqnarray} U = \frac{F(R)R-f(R)}{2\kappa^2}.\label{a5} \end{eqnarray}
We switch over to  the Einstein frame to see the real effects of dark matter in form of the scalar degree of freedom. It is possible to derive an action in the Einstein frame under the conformal transformation
\begin{eqnarray} \tilde{g}_{\mu\nu} = \Omega^2g_{\mu\nu},\label{a6}\end{eqnarray}
where $\Omega^2$ is the conformal factor and an overhead  tilde denotes the  quantities pertaining to  the Einstein frame. The corresponding  Ricci scalars in the two frames are mutually  related  as
\begin{eqnarray} R = \Omega^2(\tilde{R} + 6\tilde{\Box}\omega - 6\tilde{g}^{\mu\nu}\partial_{\mu}\omega\partial_{\nu}\omega),   \label{a7}\end{eqnarray}
where\begin{eqnarray} \omega \equiv \ln\Omega,     \partial_{\mu}\omega\equiv\frac{\partial\omega}{\partial\tilde{x}^{\mu}},    \tilde{\Box}\omega \equiv\frac{1}{\sqrt{-\tilde{g}}}\partial_{\mu}(\sqrt{-\tilde{g}}\tilde{g}^{\mu\nu}\partial_{\nu}\omega).\label{a8}\end{eqnarray}
 Thus, the action (\ref{a4}) under the conformal transformation is transformed as \cite{bc1}
 \begin{eqnarray}  {\mathcal{A} = \int d^{4}x \sqrt{-\tilde{g}}\times}\nonumber\\\left[\frac{1}{2\kappa^{2}}F\Omega^{-2}(\tilde{R} + 6\tilde{\Box}\omega - 6\tilde{g}^{\mu\nu}\partial_{\mu}\omega\partial_{\nu}\omega) - \Omega^{-4}U \right]\nonumber \\  + \mathcal{A}_{m}. \label{a9}\end{eqnarray}
The linear action in $\tilde{R}$ can be written by choosing
\begin{eqnarray} \Omega^{2} = F. \label{a10}\end{eqnarray}
We  consider a new scalar field $\phi$ defined by
\begin{eqnarray} \kappa\phi \equiv \sqrt{\frac{3}{2}}\ln F. \label{a11}\end{eqnarray}
Using the  relations (\ref{a10}) and  (\ref{a11}),  the action  in the  Einstein frame is found as
\begin{eqnarray}\mathcal{A} = \int d^{4}x \sqrt{-\tilde{g}}\left[\frac{1}{2\kappa^{2}}\tilde{R} -  \frac{1}{2}\tilde{g}^{\mu\nu}\partial_{\mu}\phi\partial_{\nu}\phi - V(\phi) \right] + \mathcal{A}_{m}, \nonumber \\  \label{a12}\end{eqnarray}
where \begin{eqnarray} V(\phi) = \frac{U}{F^{2}} = \frac{FR -f}{2\kappa^2F^2}   \label{a13}\end{eqnarray}
stands  as the potential term of the scalar degree of freedom in a general $f(R)$ model.

\section{\label{3} \textbf{Dark matter like effects of scalaron  in $f(R) = \frac{R^{1+\delta}}{R_{c}^{\delta}}$ type models}}

From the constant tangential velocity condition on the motion of a test particle in the stable orbits of the spiral galaxies, the form of $f(R)$ can be  given by $f(R)=R^{1+\delta}$, where $\delta<<1$ is related to the tangential velocity  \cite{b17}.   Therefore,   to solve the problem of dark matter, only very small deviation from general relativistic theory is required. However, we use a scalar field  approach to overcome the dark matter issue in $f(R)$ gravity as obtained through  the conformal transformation  presented in the previous section. In this approach,  it is shown that the dark matter can be scalar field particle ``scalaron". In the Einstein frame the scalar field $\phi$ is coupled with non-relativistic matter. This coupling has the relation

\begin{eqnarray} \Omega^{2} = F = e^{-2Q\kappa\phi}, \label{a14}\end{eqnarray}
where $Q$ is the strength of  coupling. Now,  from equations  (\ref{a11}) and  (\ref{a14}), $Q$ is given by
\begin{eqnarray} Q = -\frac{1}{\sqrt{6}}.  \label{a15}\end{eqnarray}
We consider the dark matter as the scalar field  arising in the  $f(R)$ model given by
 \begin{eqnarray} f(R)=\frac{R^{1+\delta}}{R_{c}^{\delta}}  \label{a16}\end{eqnarray}
 where $R_{c}$ is a constant having unit of  the  Ricci scalar $R$  and $\delta$ is a  small parameter of the model.
Differentiating equation (\ref{a16}) with respect to $R$, we obtain
\begin{eqnarray} F={(1+\delta)}\frac{R^{\delta}}{R_{c}^{\delta}}.  \label{a17}\end{eqnarray}
From equations (\ref{a14}) and (\ref{a17})  with  $Q = -\frac{1}{\sqrt{6}}$,  the Ricci scalar $R$ in terms of scalar field $\phi$ is given as
\begin{eqnarray} R = R_{c}\left[\frac{e^{\sqrt{2/3}\kappa \phi}}{1+\delta}\right]^{\frac{1}{\delta}}.  \label{a18}\end{eqnarray}
The scalar field $\phi$ remains positive for $(1+\delta) R^\delta > R^\delta_{c}$.  In case of   small  coupling  $\kappa\phi \ll 1$, $\phi=\sqrt{6} \delta /2\kappa$, for                   $R=R_c$.

The variation of the action (\ref{a12}) with respect to  $\phi$ yields  the equation of motion of the scalar field  given by
\begin{eqnarray} \tilde{\Box}\phi = V'(\phi) + \frac{\kappa}{\sqrt{6}}\tilde{T} \label{ab1}\end{eqnarray}
where $\tilde{\Box}\phi = \frac{1}{\sqrt{-\tilde{g}}}\partial_{\mu}(\sqrt{-\tilde{g}}\tilde{g}^{\mu\nu}\partial_{\nu}\phi)$, $V'(\phi)=\frac{dV}{d\phi}$ and $\tilde{T}=\tilde{g}^{\mu\nu}\tilde{T}_{\mu\nu}$. Since the traceless  electromagnetic fields will not directly couple to scalaron, as also   shown through the Lagrangian of the  massless vector fields  in the Einstein frame, the scalaron may exist   as a  dark matter particle.
Equation (\ref{ab1}) can also be written as
\begin{eqnarray} \tilde{\Box}\phi = V_{eff}'(\phi) \label{ab2}\end{eqnarray}
where $V_{eff}'(\phi)=V'(\phi) + \frac{\kappa}{\sqrt{6}}\tilde{T}$.
Using  (\ref{a13})  with  (\ref{a16})  and   (\ref{a17}) for  $ V(\phi)$ and adding it to the contribution arising  from  matter, the effective potential of the scalaron for non-relativistic matter is given by
\begin{eqnarray} V_{eff}(\phi) =\frac{\delta R_{c}}{2\kappa^2 (1+\delta)^\frac{(1+\delta)}{\delta}}e^{\sqrt{\frac{2}{3}}\frac{(1-\delta)}{\delta}\kappa\phi} + \frac{1}{4} \rho e^\frac{-4\kappa\phi}{\sqrt{6}}    \label{a19} \end{eqnarray}

%----------------------------------------------------------------------------------------------------------------------------------------------------------

To find the value of the scalar field  $\phi$ at which $V_{eff} (\phi)$ is minimum, we find out $\frac{dV_{eff}}{d\phi}$ given as
 \begin{eqnarray} V'_{eff}(\phi) =\frac{R_{c}}{\sqrt{6}\kappa}\frac{(1-\delta)}{(1+\delta)^\frac{1+\delta}{\delta}}e^{\sqrt{\frac{2}{3}}\frac{(1-\delta)}{\delta}\kappa\phi} - \frac{\kappa}{\sqrt{6}} \rho e^\frac{-4\kappa\phi}{\sqrt{6}}    \label{a20} \end{eqnarray}
Solving $V'_{eff}(\phi)=0$, we obtain  the value of  $\phi$ at the minimum of $V_{eff}(\phi)$ given by,
\begin{eqnarray} \phi_{min} = \sqrt{\frac{3}{2}}\frac{1}{\kappa}\ln\left[(1+\delta)\left(\frac{\kappa^{2}\rho}{R_{c}(1-\delta)}\right)^\frac{\delta}{1+\delta}\right]    \label{a21}\end{eqnarray}
For the calculation of the  mass of the scalar field, we have
\begin{eqnarray} V''_{eff}(\phi) =\frac{R_{c}}{3}\frac{(1-\delta)^2}{\delta(1+\delta)^\frac{1+\delta}{\delta}}e^{\sqrt{\frac{2}{3}}\frac{(1-\delta)}{\delta}\kappa\phi} + \frac{2\kappa^{2}}{3} \rho e^\frac{-4\kappa\phi}{\sqrt{6}} \label{a22}   \end{eqnarray}
which for the value of $\phi_{min}$   (\ref{a21}) becomes
\begin{eqnarray} V''_{eff}(\phi_{min})  = \frac{(1-\delta)^{\frac{2\delta}{1+\delta}}}{3\delta(1+\delta)}(R_{c})^\frac{2\delta}{\delta+1}(\kappa^2\rho)^{\frac{1-\delta}{1+\delta}}   .\label{a23} \end{eqnarray}
Thus,  mass of the scalaron  $(= V''_{eff}(\phi_{min})) $  is clearly  given by

\begin{eqnarray} m_{\phi}^2  = \frac{(1-\delta)^{\frac{2\delta}{1+\delta}}}{3\delta(1+\delta)}(R_{c})^\frac{2\delta}{\delta+1}{(\kappa^2\rho)}^{\frac{1-\delta}{1+\delta}}      .\label{a24} \end{eqnarray}

%----------------------------------------------------------------------------------------------------------------------------------------------------------

\begin{figure}[h]
\centering  \begin{center} \end{center}
\includegraphics[width=0.50\textwidth,origin=c,angle=0]{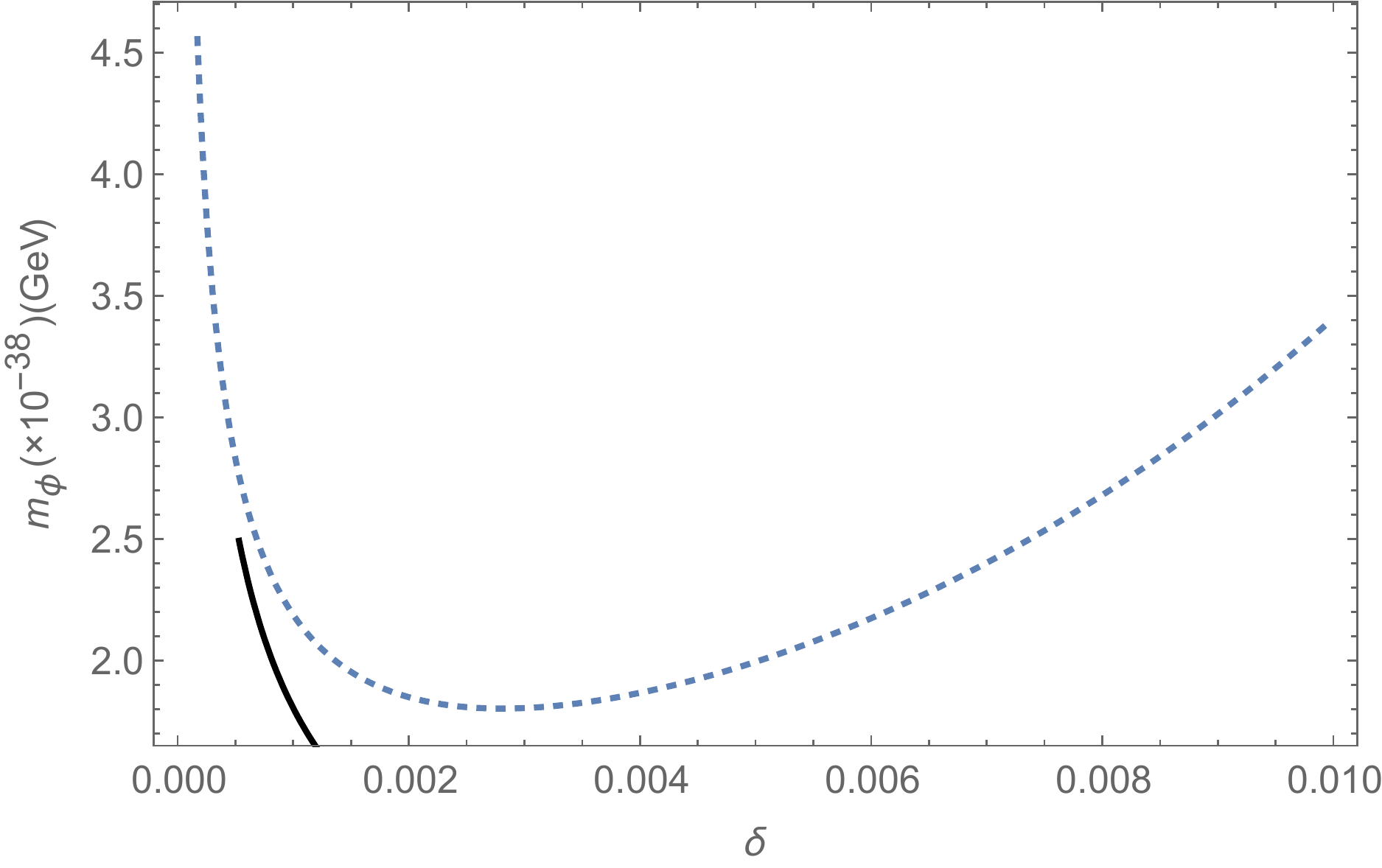}
% "\includegraphics" is very powerful; the graphicx package is already loaded
\caption{\label{fig:p1}Plot for the variation of the scalaron mass $m_{\phi}$  with parameter $\delta$.  Here, the black curve corresponds to $R_{c}=\Lambda$ (value of cosmological constant) and the dotted curve to  $R_{c}=1$.  The value of the energy density of  matter at the galactic scale is $\rho= 4\times 10^{-42} (GeV)^4$ for both curves.}\label{f1}
\end{figure}
%-----------------------------------------------------------------------------------------------------------------------------------------------------------
\begin{figure}[h]
\centering  \begin{center} \end{center}
\includegraphics[width=0.50\textwidth,origin=c,angle=0]{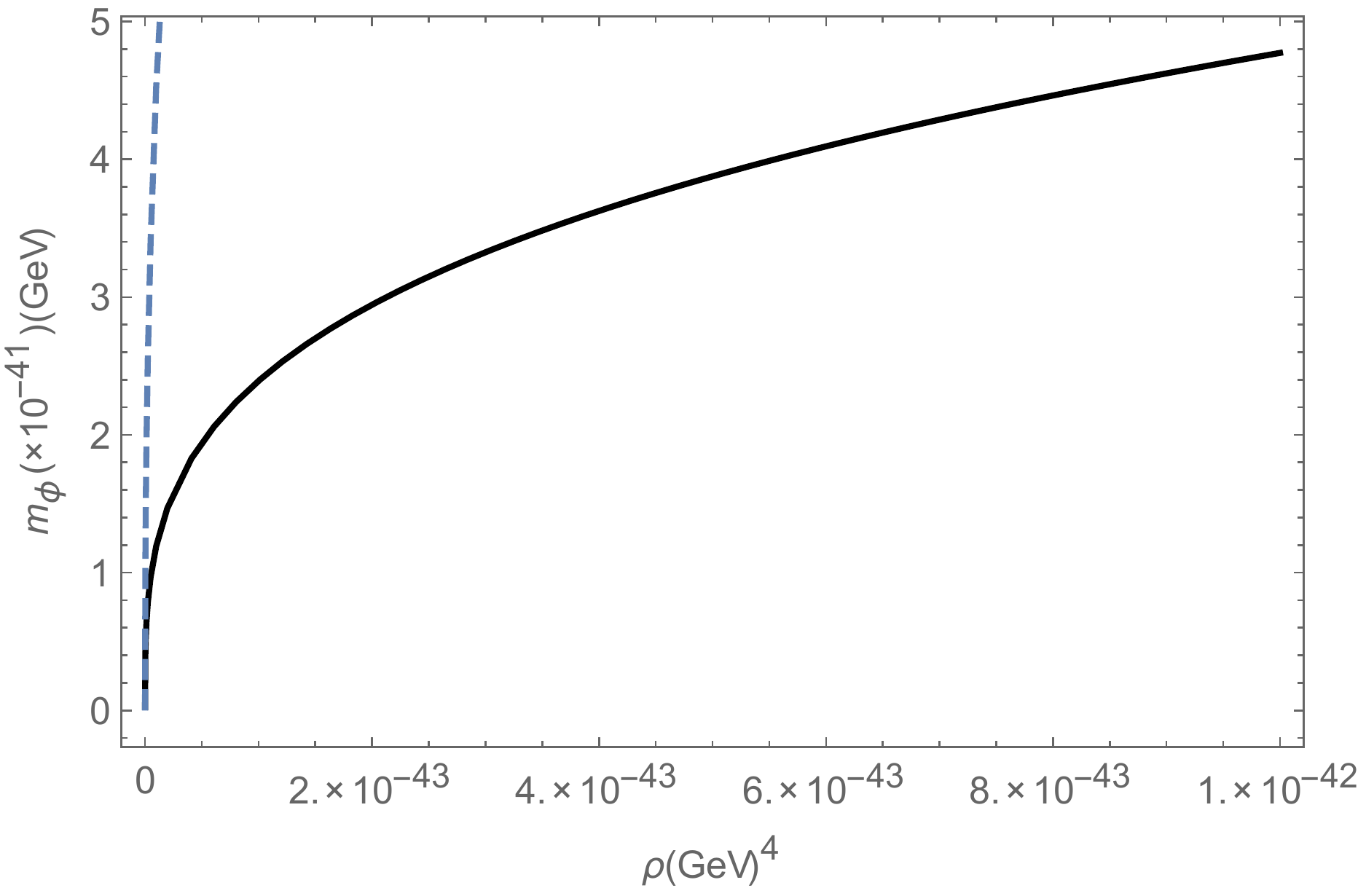}
% "\includegraphics" is very powerful; the graphicx package is already loaded
\caption{\label{fig:p2} Plot for the variation of scalaron mass $m_{\phi}$  with the energy density $\rho$ of matter. Here, black and dashed curves correspond to $\delta=0.25$ and $\delta=0.10$,  respectively, with  $R_{c}=\Lambda$.}\label{f2}
\end{figure}

%----------------------------------------------------------------------------------------------------------------------------------------------------------
\begin{figure}[h]
\centering  \begin{center} \end{center}
\includegraphics[width=0.50\textwidth,origin=c,angle=0]{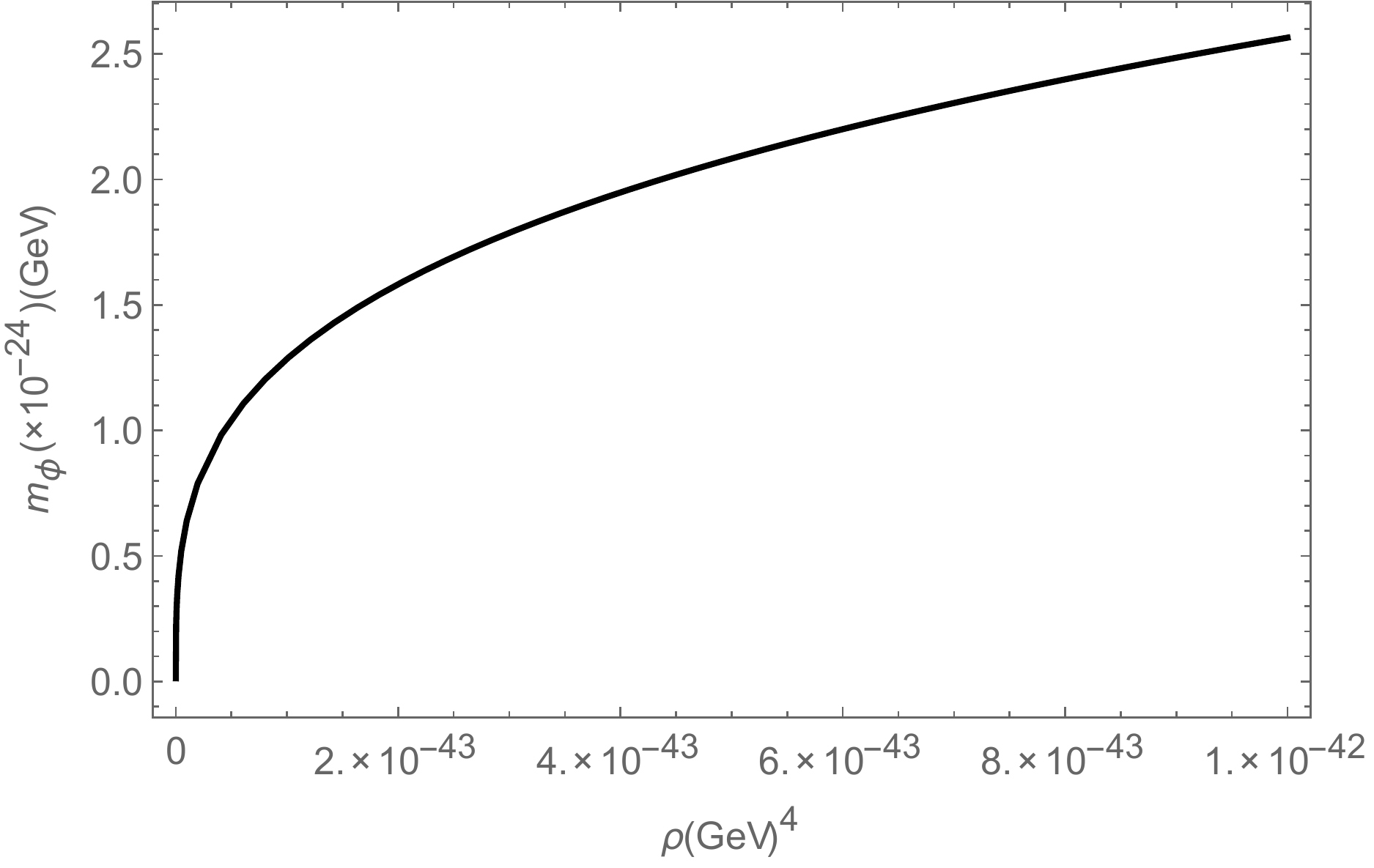}
% "\includegraphics" is very powerful; the graphicx package is already loaded
\caption{\label{fig:p3} Plot for the variation of scalaron mass $m_{\phi}$  with the energy density $\rho$ of  matter corresponding to $\delta=0.25$ and $R_{c}=1$.}\label{f3}
\end{figure}
%-----------------------------------------------------------------------------------------------------------------------------------------------------------
It becomes clear that the mass of the scalar field depends upon the energy density of  standard matter. Considering a  very small $\delta \sim 10^{-6}$,  at the electro-weak scales $\rho_{EW}\sim   (100 GeV)^4$, $m_\phi\sim 10^{-3}$  eV, while at the solar system scales $\rho_ {\bigodot}\sim 10^{19} {eV}^4$,  scalaron must be  very light as $\sim 10^{-16}$ eV.
 Fig.  \ref{f1} shows the  dependence of the scalaron  mass on  the model parameter $\delta$ over its very small values.   When $R_{c}=\Lambda$, the scalaron mass decreases with increasing $\delta$, while in  case of $R_{c}=1$, $m_{\phi}$ decreases  more sharply initially for smaller values of $\delta$ and then  increases for its  larger values  for the given  energy density of matter $\rho$  at the galactic scale. Since  $\delta$ substantially   determines the form of the $f(R)$ model, therefore, we  have the model dependent mass of the scalaron.  It tends to infinity as $\delta$ tends to zero and our model approaches the standard  general relativistic description.  From  Fig. \ref{f2}, it is obvious  that the scalar field is coupled with the standard matter  and the mass of the scalaron  becomes  large in the high curvature regions, although  the coupling is weak  around  large curvature  and  the Compton wavelength of scalaron  becomes small there. This provides a  physical constraint on the scalaron mass such that it should be able to interact as a  dark matter particle  with the standard matter in its  neighbourhood.    However,   the behaviour  of scalar field changes with the length scale due to varying energy density of matter. We notice that even a  small change of $\delta$ makes a large difference for such dependence of the mass on the  energy density background.  In particular,  Fig. \ref{f3} shows that the  behaviour of $m_{\phi}$ with respect to  energy density $\rho$ for $R_{c}=1$  is reproduced as that for $R=\Lambda $, for the same $\delta=0.25 $  even though the mass $m_{\phi}$ now gets  much higher.  From the above figures, we find that the scalaron mass $m_{\phi}$  is large  for smaller values of $\delta$ and high energy density of matter $\rho$. Since, the chameleon mechanism works in the high energy density regions,  where  the Compton wavelength of the scalaron becomes too small, therefore, the motion of the scalar field is diminished and screened out.  It is consistent with the local gravity constraints. These properties of scalaron show that  it could reproduce the effects of  dark matter. The form $f(R)=R^{1+\delta} $, where $\delta$ is of the order of $10^{-6}$ as the squared tangential velocity of a test particle  in the circular orbits   around the galactic centres,  is consistent with the local gravity constraints \cite {b17}.
It is possible to  constrain the mass of the scalaron  using the  energy density of  matter  from the large scale structure  observations and $\delta$  from the galactic rotational velocities,  with $R_c $  as a scaling parameter.   This mass can then be compared with  the mass  of  the  cold dark matter particle in the standard model,  and also with  the bounds from the decay widths of scalaron as a dark matter particle.  It is also  expected that the large  translational  velocities of the baryonic matter (such as of merging  clusters) would induce  anisotropy in the scalaron mass through the corresponding anisotropies in  $\delta$  and $R_c$.   This may cause a  directional  propagation of the scalar degree of freedom in the anisotropic,  large curvature background.     From the fact that   $\delta$,  $R_c$   and $\rho$     determine the unique  form of the effective  potential, it is clear that  we can find  the exact viable  form of the $f(R)$ model to reproduce the effects of dark matter in such high mass density regions.

\section{\label{4}\textbf{Dark matter and  the motion of the scalar field}}
 The equation of motion of the scalar field  (\ref{ab2}) can be written as
\begin{eqnarray} \frac{d^{2}\phi}{d\tilde{t}^{2}} + 3\tilde{H}\frac{d\phi}{d\tilde{t}} + V'_{eff}(\phi) = 0. \label{a25}\end{eqnarray}
where $\tilde{H}$ is the Hubble parameter in the Einstein frame.   When $\tilde{H}=0$, the energy of the system is conserved  and the oscillations are periodic.  On the other hand, when $\tilde{H}\neq0$, then it acts to  produces a  dissipative force  against  the oscillations of the scalar field. In this case,  the energy of the field  system is not conserved and the motion of the scalar field is not periodic. If the Hubble parameter $\tilde{H}$ varies slowly (adiabatically) with time during the time period $T$ of the oscillation such that
   \begin{eqnarray} \tilde{H}\ll\nu \label{a26}\end{eqnarray}
  where $\nu$ is the frequency of the oscillations,  then the rate of loss  of energy, being proportional to the Hubble parameter,  is very small and oscillations are approximately periodic.

  Previously, several authors have studied  the solution of the scalar field  in the action-angle formalism to explain the dark energy \cite{b22,b26}.  Motion of a test particle in $f(R)$ gravity has also been studied through the action angle variable formalism in our previous work  \cite{b29}.  We adopt this formalism in the present  case to derive the equation of state of scalar field using the action-angle variable $J$   which  is defined as \cite{b23,b24}
\begin{eqnarray} J= \oint p d\phi = 2\int^{\phi_{2}}_{\phi_{1}}\sqrt{2(\rho_{\phi}-V_{eff}(\phi))} d\phi \label{a27}\end{eqnarray}
where $p$ is the momentum and $\rho_{\phi}$ is the energy density of scalar field. $\phi_{1}$ and $\phi_{2}$ are the values of $\phi$ at which $V_{eff}(\phi_{1})= V_{eff}(\phi_{2})=\rho_{\phi}$.

The equation of state  $w$  of the scalar field is defined as
\begin{eqnarray}  w = -1 + \frac{J}{\rho_{\phi}}\frac{1}{dJ/d\rho_{\phi}}       .\label{a28}\end{eqnarray}
We  obtain the turning points $\phi_1$  and $\phi_2$,  and $J$ for the cases  $\delta=0.20$  and  $\delta=10^{-6}$ to  calculate  $w$.

%----------------------------------------------------------------------------------------------------------------------------------------------------------
\begin{figure}[h]\centering  \begin{center} \end{center}
\includegraphics[width=0.50\textwidth,origin=c,angle=0]{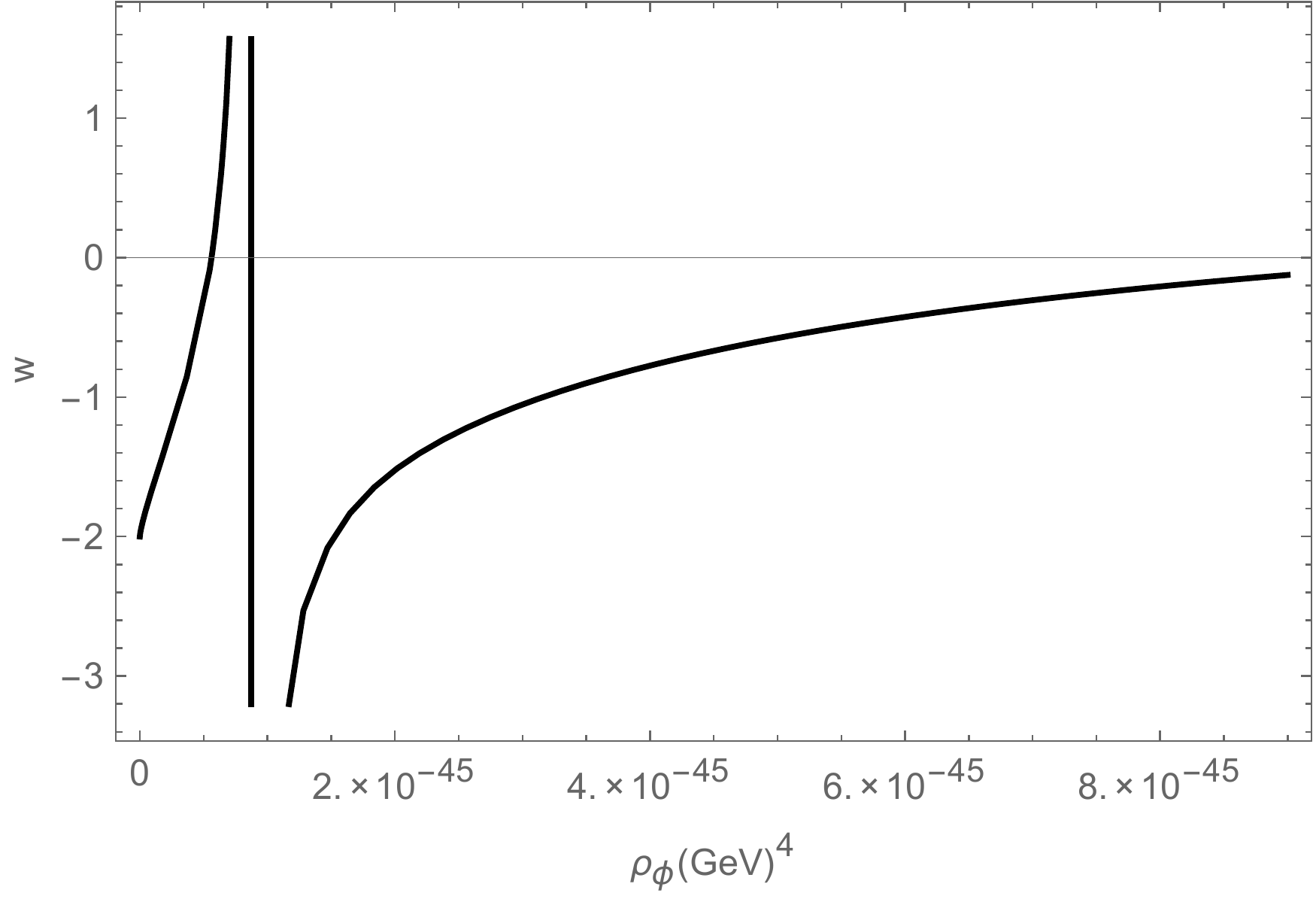}
% "\includegraphics" is very powerful; the graphicx package is already loaded
\caption{\label{fig:p4} Plot for the variation of equation of state  $w$  with the energy density $\rho_{\phi}$ of scalar field corresponding to $\delta=0.20$.}\label{f4}
\end{figure}
%-----------------------------------------------------------------------------------------------------------------------------------------------------------
%----------------------------------------------------------------------------------------------------------------------------------------------------------
\begin{figure}[h]
\centering  \begin{center} \end{center}
\includegraphics[width=0.50\textwidth,origin=c,angle=0]{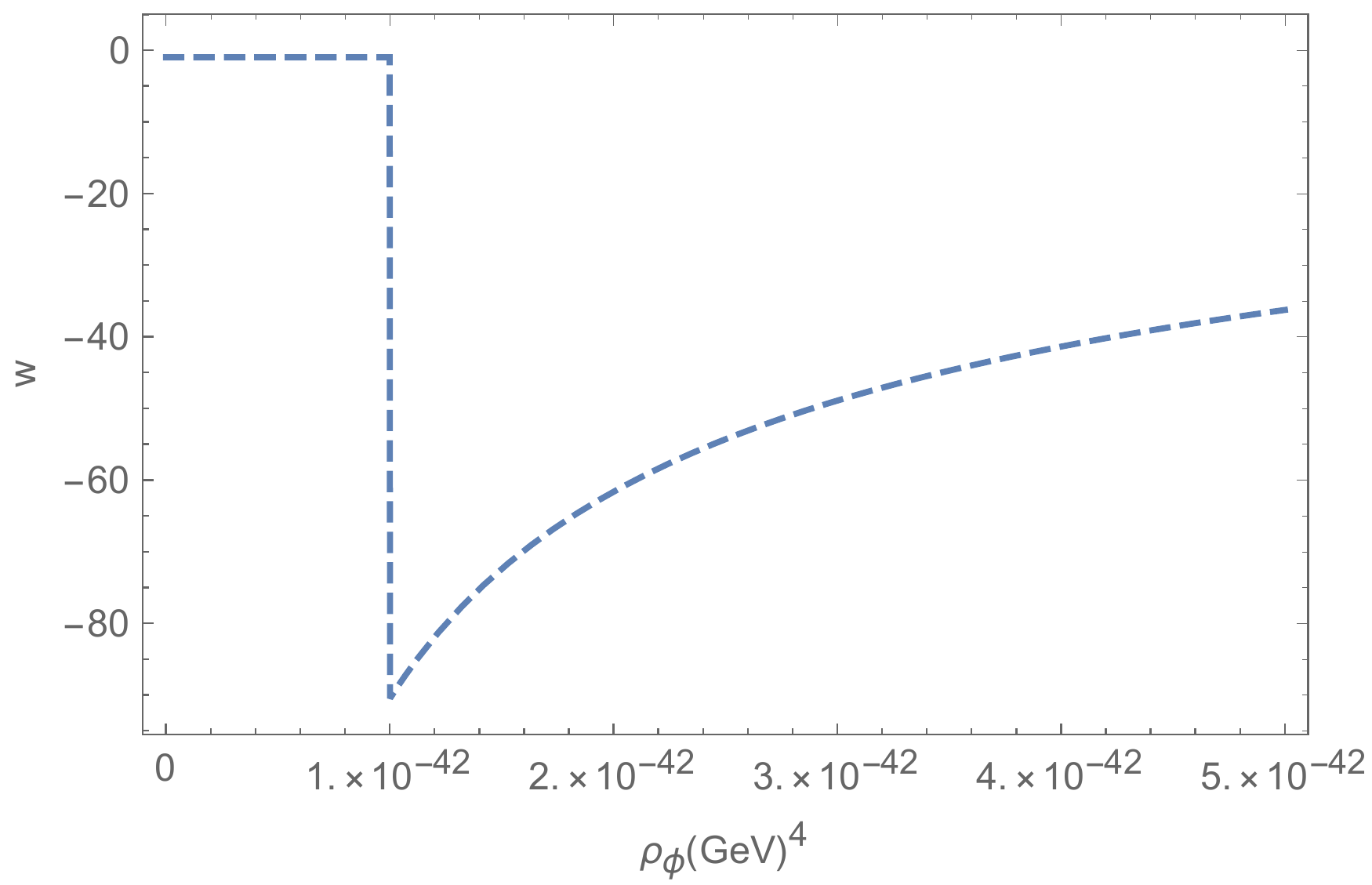}
% "\includegraphics" is very powerful; the graphicx package is already loaded
\caption{\label{fig:p5} Plot for the variation of equation of state of scalar field $w$  with the energy density $\rho_{\phi}$ of scalar field corresponding to $\delta=10^{-6}$.}\label{f5}
\end{figure}
%-----------------------------------------------------------------------------------------------------------------------------------------------------------
Figures  \ref{f4}  and  \ref{f5}  show the plots of the equation of state $w$ of the scalar field with energy density $\rho_{\phi}$ of the scalar field. Here, the energy density of the non-relativistic matter is $4\times 10^{-42} (GeV)^{4}$. It is found that $w$ is negative, zero and positive in different regions of energy density $\rho_{\phi}$. When $w=0$, then the scalar field can behave as cold dark matter and when $w=\frac{1}{3}$, it corresponds to radiation. Negative equation of state i.e. $w<0$ corresponds to the field with negative pressure.
From Fig. \ref{f4},  we find that initially, when energy density is small, equation of state is negative. In such case, with an upper bound, the  scalar field behaves as dark energy.  Further,   when $\rho_{\phi}$ increases,  equation of state rises to  zero and it  acts like cold dark matter. When $\rho_{\phi}$ increases further, equation of state $w$ becomes positive.

Fig. \ref{f5}  shows the behaviour of equation of state $w$  with energy density of the scalar field $\rho_{\phi}$ with  $\delta=10^{-6}$ i.e. a very small deviation from the  general  relativity. Here, $w=-1$, for a certain range of energy density $\rho_{\phi}$ of scalar field. At a certain value of energy density $\rho_{\phi}$, equation of state $w$  decreases drastically and after that $w$ approaches to $-1$, when $\rho_{\phi}$ increases further.
\section{\label{5}\textbf{Summary and Discussion}}
We investigated here the possibility of explaining the dark matter problem  in the framework of  $f(R)$ gravity. In the Einstein frame,  scalaron appears  as the dark matter particle that does  not directly  couple to the traceless electromagnetic fields and  remains  dark with respect to such interaction. Further,  being a fluctuation in a background of scalar field $\phi_{min}$ where $V_{eff}(\phi)$ is minimum,   this does not need to  add any  new matter component to the particle inventory  of the standard model.  However, the effective scalaron potential  is  influenced by the energy density of the  standard matter in the background.  In our model,  with  $f(R)=\frac{R^{ 1+\delta }}{R_{c}^{\delta}}$,  the mass of the scalaron almost linearly depends upon the energy density of  standard matter.  Therefore, it does not  vary  much in large curvature regions  in comparison to the other models (like the  Starobinsky model \cite{bc2}) in which there is a power law dependence. This behaviour is suitable for the dark matter phenomena under screening conditions. In our model, the mass of the scalaron can be constrained in two ways  using particle physics as well as the  cosmological observations. One way is to find  the decay widths of the scalaron and check the bounds on  its  mass.  The other way is to obtain it   directly  from the observational data of the background  matter energy density and the  model parameters  $\delta$ and   $R_c$  (from the galactic rotation and lensing  etc.).  This mass must  then also  be compared  with  the theoretical  constraints  obtained  in the $\Lambda$CDM model.   We have not found  any conditions  on the amount of the  total scalaron  content, or its time evolution,   or  the scale dependence of the $\delta$  and  $R_c$    i.e. the deviation from the Einstein's general relativistic theory at different scales,  especially     in  the  background of  violent  mergers of clusters.  In  such cases, the role of $\delta$ and $R_c$  both is unclear, and may profoundly indicate a   solution with anisotropic propagation of scalaron in response to the direction and magnitude  of  velocities  of standard baryonic  matter.  This  approach   may lead  to solve the puzzles of  the  offset of  dark matter from the intra-cluster baryonic hot gases, as marked in the  Bullet cluster (1E0657-56) \cite{bc3},   Abell  520 system \cite{bc4}, MACS \cite {bc5} etc.     Clearly,  lensing must  be a good tool to investigate  such scale  dependence and anisotropic propagation of the scalar degree of freedom in the large curvature  background,  with additional tools  to check  whether the large scale (low curvature) behaviour of the scalaron may produce the effects of dark energy causing  acceleration in recent epochs.    In general,  it seems unlikely that $\delta$ and $R_c$  must remain constant and isotropic  over all scales and also throughout the time evolution of the universe. In view of the action-angle formalism,  the variation of equation of state of the field with its energy density opens  a way to connect the dark energy to dark matter in a single perspective.   We will attempt to study  these broader issues  in our future work.

\section*{\textbf{Acknowledgments}}
 MMV thanks the Kavli IPMU, Tokyo, especially  Hitoshi Murayama,  Misao Sasaki and  Masahiro Takada  for hosting the visit  and useful discussions.  Both authors thank  IUCAA, Pune  for the facilities where  a part of the present work was completed under the associateship programme.


\begin{thebibliography}{00}
\bibitem{b1} A. G. Riess et al., (Supernova Search Team Collaboration), Astron.  J. \textbf{116}, 1009 (1998).
\bibitem{b2} B. Schmidt et al.,  Astrophys.  J.  \textbf{507}, 46 (1998).
\bibitem{b3} S. Perlmutter et al.,  Astrophys.  J. \textbf{517}, 565 (1999).
\bibitem{b4} D. N. Spergel et al.,  Astrophys.  J. Suppl. \textbf{148}, 97 (2003).
\bibitem{b5} S. M. Carroll, Living Rev. Relativ.(2001)4:1.
\bibitem{b6} P. J. E. Peebles and B.  Ratra,  Rev. Mod. Phys. \textbf {75}, 559-606 (2003).
\bibitem{b7} R. R. Caldwell, R. Dave, and P. J. Steinhardt, Phys. Rev. Lett. \textbf{80},   1582 (1998).
\bibitem{b8} S. Capozziello, Int. J. Mod. Phys. D \textbf{11}, 483 (2002).
\bibitem{b9} T. Chiba, T. Okabe, and M. Yamaguchi, Phys. Rev. D \textbf {62}, 023511 (2000).
\bibitem{b10} R. R. Caldwell,  Phys. Lett. B \textbf{545}, 23 (2002).
\bibitem{b11} M. Kunz and D. Sapone, Phys. Rev. Lett. \textbf{98}, 121301 (2007).
\bibitem{b12} L. Amendola, R. Gannouji, D. Polarski and S. Tsujikawa, Phys. Rev. D \textbf{75}, 083504 (2007).
\bibitem{b13} V. Sahni and Y. Shtanov,  JCAP \textbf{0311}, 014 (2003).
\bibitem{b14} A. A. Starobinski, Phys. Lett. B \textbf{91}, 99(1980).
\bibitem{c1}  S. Nojiri and S. D. Odintsov,  Phys. Rept. \textbf{505}, 59 (2011); S. Nojiri and S. D. Odintsov  and  V . K. Oikonomou,   Phys. Rept. \textbf{692}, 1, (2017).
\bibitem{b15} V. C. Rubin, W. K. Ford and N. Thonnard, Astrophys. J. \textbf{238}, 471 (1980).
\bibitem{b16} A. Borriello and P. Salucci, Mon. Not. R. Astron. Soc. \textbf{323}, 285 (2001).
\bibitem{c2}  G. Jungman, M. Kamionkowski and K. Griest, Phys. Rep. \textbf{267}, 195 (1996).
\bibitem{b17} C. G. Boehmer, T. Harko and F. S. N. Lobo, Astropart. Phys. \textbf{29}, 386 (2008).
\bibitem{b171} C. Corda, H. J.  Mosquera Cuesta and R. L. Gomez,  Astropart.  Phys. \textbf{35},  362  (2012).
\bibitem{b18} T. Katsuragawa and S. Matsuzaki, Phys. Rev. D \textbf{95}, 044040 (2017); Phys. Rev. D. \textbf{97}, 6, 064037  (2018).
\bibitem{b19} R. Zaregonbadi, M. Farhoudi and N. Riazi, Phys. Rev. D \textbf{94}, 0840052 (2016).
\bibitem{b20} V. Sahni and L. Wang, Phys. Rev. D \textbf{62}, 103517 (2000).
\bibitem{b21} S. S. Mishra, V. Sahni and Y. Shtanov JCAP \textbf{06}, 045 (2017).
\bibitem{b25} M. S. Turner, Phys. Rev. D \textbf{28}, 1243 (1983).
\bibitem{b28} M. M. Verma and B. K. Yadav, Int. J. Mod. Phys. D \textbf{27}, 1850002 (2018).
\bibitem{b22} E. Masso, F. Rota and G. Zsembinszki, Phys. Rev. D \textbf{72}, 084007 (2005).
\bibitem{b26} S. Dutta and R. J. Scherrer, Phys. Rev. D \textbf{78}, 083512 (2008).
\bibitem{b29} B. K. Yadav and M. M. Verma, Int. J. Mod. Phys. D \textbf{26}, 1750183 (2017).
\bibitem{b23} H. Goldstein, \emph{Classical Mechanics} (Addison Wesley, Reading MA, 1980) 2nd ed.
\bibitem{b24} L. D. Landau and E. M. Lifshitz, \emph{Mechanics}  (Butterworth-Heinemann, Oxford, UK, $1976$).
\bibitem{bc1} A. D. Felice and S. Tsuzikawa, Living Rev. Relativity \textbf{13}, 3, (2010).
 \bibitem{bc2} A.  A. Starobinsky,  JETP Lett. \textbf{86},  157   (2007).
\bibitem{bc3}  M.  Markevitch  et al.,  Astrophys. J.  \textbf{606},  819  (2004).
\bibitem{bc4}   A.  Mahdavi    et al.,  Astrophys. J.  \textbf{668},   806  (2007).
\bibitem{bc5}  M.  Bradac  et al.,   Astrophys. J.  \textbf{687},  959  (2008).

\end{thebibliography}
\end{document}